\documentclass{kluwer}
\usepackage{graphicx}

\newcommand{\A}[1]{\mathbf{#1}}

\newcommand{\DS}{\displaystyle}
\newcommand{\pd}[2]{\frac{\partial #1}{\partial #2}}
\newcommand{\HALF}{\frac{1}{2}}

\begin{document}
\begin{article}

\begin{opening}
\title{Relativistic MHD Simulations of Jets with Toroidal Magnetic Fields.}
\author{Andrea \surname{Mignone}\email{mignone@to.astro.it}}
\author{Silvano \surname{Massaglia}}
\author{Gianluigi \surname{Bodo}}
\institute{Dipartimento di Fisica Generale dell'Universit\`a,
Via Pietro Giuria 1, I-10125 Torino, Italy \\
Osservatorio Astronomico di Torino, Viale Osservatorio 20, I-10025
Pino Torinese, Italy\\}

\runningauthor{Mignone et al.} 
\runningtitle{RMHD Simulations} 
\date{July 30, 2005}

\begin{abstract}
 This paper presents an application of the recent relativistic HLLC 
 approximate Riemann solver by Mignone \& Bodo to 
 magnetized flows with vanishing normal component of the magnetic field.
 The numerical scheme is validated in two dimensions by investigating   
 the propagation of axisymmetric jets with toroidal magnetic fields.
 The selected jet models show that the HLLC solver yields sharper resolution
 of contact and shear waves and better convergence properties 
 over the traditional HLL approach. 
\end{abstract}

\end{opening}

\section{Introduction}
%
%
%

Jets are one of the fundamental components of radio-loud 
active galactic nuclei (AGN).
Although different in detail, they all share the property 
of being relativistic on the parsec scale and of carrying 
magnetic fields that may be dynamically important.
Observations of their non-thermal synchrotron radiation, in fact,
require the presence of a magnetic field and the high degree of polarization
observed indicates the presence of some large scale structure in the
field.
 
Theoretical investigations require,
therefore, a relativistic MHD description of such objects.
In this effort, recent numerical simulations \cite{K99b,LAAM05}
have turned out as efficient investigation tools in understanding 
many aspects of the jet evolution.
This, in turn, justifies recent and new efforts to extend 
existing gas-dynamical Godunov-type codes to the relativistic regime,
see \inlinecite{K99a}, \inlinecite{Balsara01}, \inlinecite{dZBL03},  and 
the extensive review by \inlinecite{MM03}.
In this perspective, we extend the recently 
derived relativistic hydrodynamical HLLC Riemann solver \cite{MB05} to 
the magnetic case, with vanishing normal component of the magnetic field.
The novel numerical scheme is then applied to the  
propagation of relativistic jets in presence of toroidal
magnetic fields.
The paper is organized as follows: in \S\ref{sec:equations}
we give the relevant relativistic MHD equations, in 
\S\ref{sec:hllc} we briefly review the approximate Riemann
solver. Finally, in \S\ref{sec:tests} we discuss the astrophysical
jet applications.

\section{RMHD Equations}\label{sec:equations}
%
%
%

The motion of a magnetized, ideal relativistic fluid is 
governed by the system of conservation laws \cite{Anile89}
\begin{equation}
 \pd{\A{U}}{t} + \sum_d \pd{\A{F}^d}{x^d} = 0 \; ,
\label{eq:rmhd_eq}
\end{equation}
where $d = x,y,z$ and $\A{U} = (D, m_k, B_k, E)$
is the vector of conservative variables with components 
given, respectively, by the laboratory 
density $D$, the three components of momentum $m_k$ and 
magnetic field $B_k$ ($k=x,y,z$) and the total energy density $E$.
$\A{F}^k$ are the fluxes along the $x^k$ direction;
for $k=x$ one has
\begin{equation}\label{eq:fluxes}
  \A{F}^x = \left[
  Dv_x, \, 
  m_kv_x - B_x\left(\frac{B_k}{\gamma^2} + (\A{v}\cdot\A{B})v^k\right) + p\delta_{xk},\,
  B_kv_x - B_xv_k, \,
  m_x\right]\,, \quad
\end{equation}
where $p = p_g + |\A{B}|^2/(2\gamma^2) + (\A{v}\cdot\A{B})^2/2$ is the total
pressure, $p_g$ is the thermal (gas) pressure and $\A{v}\equiv (v_x,v_y,v_z)$ 
is the fluid three-velocity. The Lorentz factor is denoted with $\gamma$.
Expressions for $\A{F}^y(\A{U})$ and $\A{F}^z(\A{U})$ follow by 
cyclic permutations of the indexes.

Introducing the primitive flow variables 
$\A{V} = (\rho, \A{v}, p_g, \A{B})$, one has
\begin{eqnarray} 
 D   & = & \rho\gamma   \;,  \label{eq:cons_var_D}\\ \noalign{\medskip}
 m_k & = & (\rho h\gamma^2 + |\A{B}|^2)v_k - (\A{v}\cdot\A{B})B_k  \; , 
          \label{eq:cons_var_m}\\ \noalign{\medskip}
 E   & = & \DS \rho h\gamma^2  - p_g  
            + \frac{|\A{B}|^2}{2} + \frac{|\A{v}|^2|\A{B}|^2 - (\A{v}\cdot\A{B})^2}{2}
             \label{eq:cons_var_E}\;,
\end{eqnarray}
where $\rho$ and  $h$ are, respectively, the rest-mass density and 
specific enthalpy of the fluid. 
An additional equation of state is necessary to close 
the system (\ref{eq:rmhd_eq}). Throughout the following we will 
assume a constant $\Gamma$-law, with specific enthalpy given
by
\begin{equation}\label{eq:eos}
  h = 1 + \frac{\Gamma}{\Gamma - 1}\frac{p_g}{\rho} \;,
\end{equation}
where $\Gamma$ is the constant specific heat ratio.

\subsection{Recovering primitive variables}
\label{sec:contoprim}
%
%

Godunov-type codes \cite{Godunov59} are based on a conservative formulation
where laboratory density, momentum, energy and magnetic fields 
are evolved in time.
On the other hand, primitive variables, 
$\A{V} = (\rho, \A{v}, p_g, \A{B})$, are required for the 
computation of the fluxes (eq. \ref{eq:fluxes}) and more 
convenient for interpolation purposes.

This task requires the solution of the nonlinear equation 
\begin{equation}
f(W) \equiv W - p_g + \left(1 - \frac{1}{2{\gamma}^2}\right)|\A{B}|^2 - E 
            - \frac{S^2}{2W^2} = 0 \;
\label{eq:press_fun}\end{equation}
where $W = \rho h\gamma^2$, $S = \A{m}\cdot\A{B}$. 
Equation (\ref{eq:press_fun}) follows 
directly from equation (\ref{eq:cons_var_E}) together with 
(\ref{eq:cons_var_m}).
Since, at the beginning of a time step $E$, $\A{B}$ and $\A{m}$ are
known quantities, both $\gamma$ and $p_g$ may be expressed in terms 
of $W$ alone from 
\begin{equation}\label{eq:c2p_m}
|\A{m}|^2 =  \left(W + |\A{B}|^2\right)^2
           \left(1 - \frac{1}{\gamma^2}\right) -
           \frac{S^2}{W^2} \left(|\A{B}|^2 + 2W\right) \;,
\end{equation}
{\bf and from the equation of state (\ref{eq:eos}) using $\rho = D/\gamma$}.
Equation (\ref{eq:press_fun}) can be solved by any standard root 
finding algorithm.
Once $W$ has been found, the Lorentz factor is easily found 
from (\ref{eq:c2p_m}), the thermal pressure from 
(\ref{eq:eos}) and velocities are found by inverting 
equation (\ref{eq:cons_var_m}): 
\begin{equation}
  v_k = \frac{1}{W + |\A{B}|^2}\left(m_k + \frac{S}{W}B_k\right)
\end{equation}
Finally, equation (\ref{eq:cons_var_D}) is used to determine
the proper density $\rho$.

\section{A relativistic HLLC Riemann solver}\label{sec:hllc}
%
%
%

In the traditional Godunov approach \cite{Godunov59}, 
the conservation laws (\ref{eq:rmhd_eq}) are advanced in time 
by solving, at each zone interface $x_{i+\HALF}$, a Riemann problem with initial 
condition\footnote{In what follows we take the $x$ axis as the normal
direction.}
\begin{equation}\label{eq:riemann}
 \A{U}(x,0) = \left\{\begin{array}{ccc}
   \A{U}_{L,i+\HALF} & \quad \textrm{if} \; & x < x_{i+\HALF} \,, \\ \noalign{\medskip}
   \A{U}_{R,i+\HALF} & \quad \textrm{if} \; & x > x_{i+\HALF} \,, \\ \noalign{\medskip}
\end{array}\right.
\end{equation}
where $\A{U}_{L,i+\HALF}$ and $\A{U}_{R,i+\HALF}$ are the left
and right edge values at zone interfaces.
The exact solution to the Riemann problem for the relativistic MHD equations 
has been recently derived by \inlinecite{GR05}. As in the classical
case, a seven-wave pattern emerges from the decay of an initial discontinuity 
separating two constant states. The resulting structure can 
be found by iterative techniques and it can be quite involved.

For the present purposes, however, we will be concerned with the 
special case where the component of magnetic field normal to a 
zone interface vanishes.
Under this condition, a degeneracy occurs where the tangential, 
Alfv{\'e}n and slow waves all propagate at the speed of the 
fluid and the solution simplifies to a three-wave pattern.
In this special situation, the relativistic HLLC approximate 
Riemann solver originally derived for the gas-dynamical equations
by \inlinecite{TSS94} and recently extended to the relativistic
regime by \opencite{MB05} (paper I henceforth), 
can still be applied with minor modifications.
This statement stems from the fact that both state and
flux vectors on either side of the tangential discontinuity 
(denoted with the $*$ superscript) retain the same form as in the 
non-magnetized case:
\begin{subequation}
 \begin{equation} \label{eq:hllc_state}
  \A{U}^*_{L,R} = \left(D^*,\, 
                  m^*_x,\,
                  m^*_y,\, 
                  m^*_z,\, 
                  B^*_y,\, 
                  B^*_z,\,
                  E^*\right)_{L,R}\,,
 \end{equation}
 \begin{equation}\label{eq:hllc_flux}
 \A{F}^*_{L,R} = \left(D^*v^*_x,\, 
                       m^*_xv^*_x + p^*,\, 
                       m^*_yv^*_x,\,
                       m^*_zv^*_x,\,
                       B^*_yv^*_x,\, 
                       B^*_zv^*_x,\, 
                       m^*_x\right)_{L,R}\,,
 \end{equation}
\end{subequation}
where we can still set $m_x^* = (E^* + p^*)v_x^*$.
Total pressure and normal velocity 
are continuous across the middle wave, i.e. $p^*_L = p^*_R$,
$v^*_{x,L} = v^*_{x,R}$.
The method of solution is thus entirely analogous to 
the strategy shown in paper I and we will not repeat it here.
The two additional components related to the
presence of transverse magnetic field are decoupled 
from the rest and can be solved without additional complications.
Once $v^*_x$ has been found, in fact, we use the jump
conditions to get
\begin{equation}
 B^*_{y,z} =  \DS \frac{\lambda - v_x}{\lambda - v_x^*}\, B_{y,z}  
\end{equation}
from the pre-shock states.

Finally, we need an estimate for the outermost right and left-going
fast wave speeds, $\lambda_R$ and $\lambda_L$.
These two signal velocities may be found by 
exploiting the characteristic decomposition of the RMHD equation,
extensively analyzed by \inlinecite{AP87} and \inlinecite{Anile89}.
In the general case, the fast and slow magneto-sonic waves satisfy a 
quartic equation solvable by means of numerical and analytical 
methods, e.g. \inlinecite{dZBL03}. 
For vanishing normal component of 
the magnetic field, however, the two roots corresponding to the 
degenerate slow waves have the trivial solution $\lambda = v_x$, whereas
the fast speeds satisfy the following quadratic:
\begin{equation}\label{eq:eigenvalues_eq}
 (\lambda - v_x)^2 = \sigma_s(1 - \lambda^2) \,,
\end{equation}
with 
\begin{equation}
 \sigma_s = \frac{B^2/\gamma^2 + \rho hc_s^2 + (\A{v}\cdot\A{B})^2(1 - c_s^2)}
                 {\rho h (1-c_s^2)\gamma^2}
\end{equation}
where $c_s = \sqrt{\Gamma p/(\rho h)}$ is the speed of sound.
The solutions of (\ref{eq:eigenvalues_eq}) are used to provide suitable
guesses for the outermost signal velocities in our HLLC Riemann
solver. As in paper I, we set
\begin{equation}
  \lambda_L = \min\big(\lambda_-(\A{V}_L), \lambda_-(\A{V}_R)\big) \;,
  \quad
  \lambda_R = \max\big(\lambda_+(\A{V}_L), \lambda_+(\A{V}_R)\big) \;,
\end{equation}
where $\lambda_{-}$ ($\lambda_+$) is the smallest (biggest) fast wave speed.


\section{Code Validation}\label{sec:tests}
%
%
%

Spatial and temporal discretization of equations (\ref{eq:rmhd_eq})
is based on the second-order Hancock predictor step already 
described in paper I. For the sake of conciseness we will not repeat it
here.

\subsection{A Shock-Tube Example}
%
%

As an illustrative test, we consider an initial discontinuity 
separating two constant left and right states given by
\begin{equation}
 \big(\rho, p_g, v_x, v_y, v_z, B_y, B_z\big) = 
 \left\{\begin{array}{cccccccc}
  \big(1  ,& 30,& 0,& 0,& 0,& 20,& 0\big)_L & \textrm{for}\quad x < 0.5\,,\\ \noalign{\medskip}
  \big(0.1,&  1,& 0,& 0,& 0,&  0,& 0\big)_R & \textrm{for}\quad x > 0.5\,.\\ \noalign{\medskip}
\end{array}\right.
\end{equation}
\begin{figure}
 \centerline{\includegraphics[width=115mm]{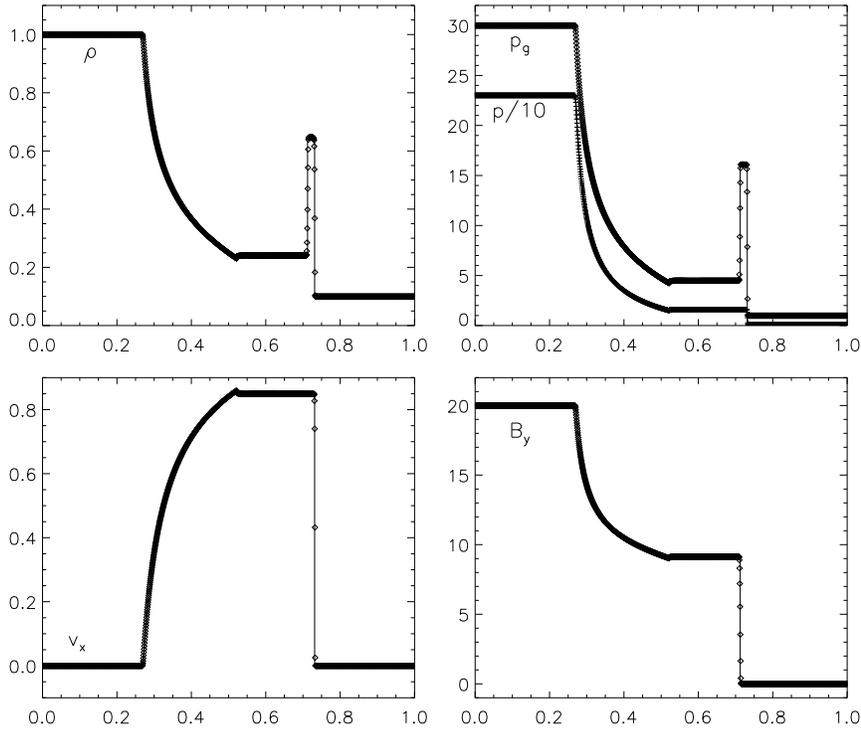}}
 \caption{Computed profiles {\bf (symbols) and exact solution (solid line)} 
          for the first test problem. 
          The second-order {\bf scheme with the} HLLC Rieman solver has 
          been employed on $1600$ zones. The final integration time is $t=0.25$;
          the Courant number is $0.5$. 
          The solution is comprised of a left-going fast rarefaction,
          a right-going tangential and shock waves moving close to
          each other.}
 \label{fig:sod1}
\end{figure}
The domain is the interval $[0,1]$ covered with $1600$ uniform computational
cells. The ideal equation of state (\ref{eq:eos}) is used with $\Gamma = 4/3$.
Figure (\ref{fig:sod1}) compares the result computed with the HLLC solver
with the analytical solution {\bf (available from \inlinecite{GR05})}, 
at the final time $t=0.25$.
The breakup of the discontinuity results in a left-going fast rarefaction wave
and a thin shell of high density material bounded by a tangential discontinuity and a 
gasdynamical shock propagating to the right.  
The total pressure and magnetic field are continuous through the contact 
and the shock wave, respectively. 
All discontinuities are correctly captured, with satisfactory resolutions:
the tangential discontinuity and the shock smear out over $\sim 5\div 6$ and $4\div 5$ 
zones, respectively. 
Small overshoots appear at the tail of the rarefaction fan. 
A similar, perhaps more pronounced behavior is shown in the results by 
\inlinecite{K99a} who used a linearized, Roe-like Riemann solver.
The relative error in the thin shell is $\lesssim 1\%$ for both 
pressure and density. 

\subsection{Axisymmetric Jet with Toroidal Magnetic Field}
           \label{sec:jet}
%
%
%

We now turn our attention to the propagation of a relativistic 
magnetized jet in cylindrical coordinates $(r,z)$.
For the sake of comparison, we have implemented three models
already discussed in literature.
The first setup, denoted with $S1$, is model $A$ of 
\inlinecite{K99b} (K99 henceforth), the second and third setups,
$S2$ and $S3$, are taken from models $C2\_10/3$ and 
$C2\_1/20$ of \inlinecite{LAAM05} (L05 henceforth), respectively. 
 
\begin{table}[t]
\begin{tabular}{cccccccc}\hline
Model & $\rho_j$  & $\rho_e$ & $v_j$         & $r_m$  & $\bar{\beta}$ & $\Gamma$ & $\sigma$\\ \hline\hline
$S1$  & $1$       & $10^3$   & $\sqrt{0.99}$ & $0.37$ &    $1$       & $4/3$     & $0.34$\\  \hline
$S2$  & $10^{-2}$ & $1$      & $0.99$        & $0.6$  &    $10/3$    & $5/3$     & $0.11$ \\ \hline
$S3$  & $10^{-2}$ & $1$      & $0.99  $      & $0.6$  &    $1/20$    & $5/3$     & $0.0017$\\ \hline
\end{tabular}
\caption{Relevant parameters for models $S1$, $S2$ and $S3$ described in 
         the text. From left to right: jet density, ambient medium density,
         beam velocity, magnetization radius, average magnetization, specific
         heat ratio and magnetic to rest mass energy ratio.}
\label{tab:ic}
\end{table}

In all models, the static external medium has uniform density 
$\rho_e$ and zero magnetic field.
At the inlet, $r \le 1$ and $z=0$, a supersonic beam with uniform velocity $v_j$ 
(Lorentz factor $\gamma_j$) and density $\rho_j$ is injected with a 
toroidal magnetic field obeying
\begin{equation}
 B_\phi(r) = \left\{\begin{array}{ccc}
  \gamma_j b_m r/r_m & \quad \textrm{if} \quad & r < r_m     \,,\\ \noalign{\medskip}
  \gamma_j b_m r_m/r & \quad \textrm{if} \quad & r_m < r < 1 \,,\\ \noalign{\medskip}
\end{array}\right.
\end{equation}
where $r_m$ is the magnetization radius of the beam.
The ideal equation of state (\ref{eq:eos}) is used in all calculations.
In model $S1$, we set $b_m = 1$ and the thermal pressure inside the beam, $p_j$, is 
prescribed from the condition of hydromagnetic equilibrium,
\begin{equation}
 p_j(r) = \left\{\begin{array}{ccc}
 p_e\left[\alpha + \frac{2}{\beta}\left(1 - (r/r_m)^2\right)\right] & \quad \textrm{if} \quad & r < r_m     \;, \\ \noalign{\medskip}
 \alpha p_e                                                         & \quad \textrm{if} \quad & r_m < r < 1 \;,\\ \noalign{\medskip}
\end{array}\right.
\end{equation}
where $\alpha = 1 - r_m^2/\beta$, $\beta = 0.34$ and 
$p_e = \beta b_m^2/2$ is the thermal pressure of the ambient medium. 
In models $S2$ and $S3$ $b_m$ is computed
from the average magnetization parameter $\bar{\beta}$: 
\begin{equation}
 b_m = \sqrt{\frac{4p_b\bar{\beta}}{r_m^2(1 - 4\log r_m)}}\,,
\end{equation}
and the thermal pressure follows directly from the
definition of the Mach number, i.e.
\begin{equation}
 p_j = \frac{\rho_j v_j^2(\Gamma-1)}{\Gamma(\Gamma-1) M_j^2 - \Gamma v_j^2} \,,
\end{equation}
with $M_j = 6$. The numerical values of the relevant parameters 
are given in Table \ref{tab:ic}.
The simulations are performed on the physical domain $r\in[0,10.5]$,
$z\in[0,50]$ using $20$ points per jet radius.  
The total grid size is $210\times 1000$.

For each model, we have carried two set of simulations: 
the first one employing the HLLC approximate Riemann solver described 
in \S\ref{sec:hllc} and the second one adopting the HLL 
scheme, also used in L05.

\begin{figure}
 \centerline{\includegraphics[width=115mm]{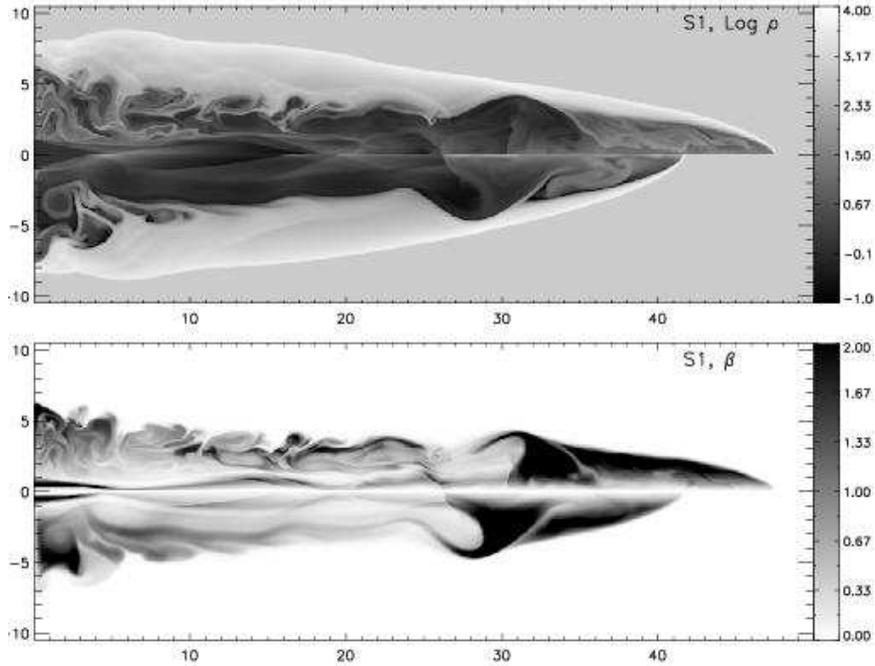}}
 \caption{Model S1 at $t = 100$. The top and bottom panels show,
          respectively, the density logarithms and magnetization 
          parameter $\beta$ for the HLLC  
          (upper half of each panel) and HLL (lower half of 
          each panel) runs. Notice that the two panels use opposite 
          color gradients. The grid resolution is $20$ zones per jet
          beam, giving a total grid size of $210$ (in $r$) and 
          $1000$ zones (in $z$). Integration has been carried with CFL = 0.4
          and the second-order Hancock scheme outlined in paper I.}
 \label{fig:s1jet}
\end{figure}
\begin{figure}
 \centerline{\includegraphics[width=115mm]{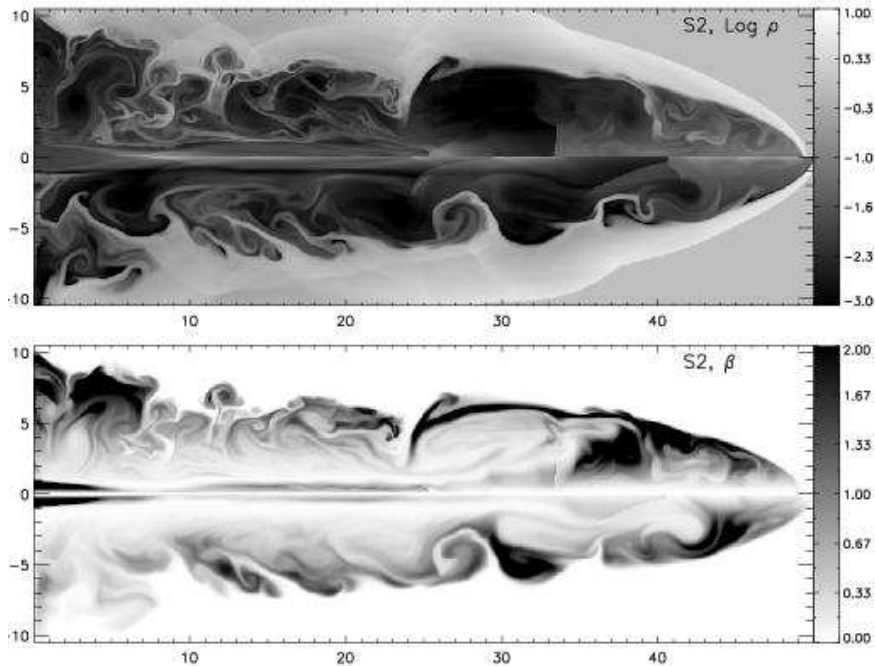}}
 \caption{Model S2 at $t = 126$. The top panel shows the density logarithm
          in the HLLC (upper half) and HLL (lower half) runs.
          The bottom panel shows the magnetization parameter 
          in the HLLC (upper half) and HLL (lower half) runs.
          The final integration time is the same one as in L05.}
 \label{fig:s2jet}
\end{figure}
Models $S1$ and $S2$ are characterized by a high Poynting flux
$B^2_\phi v$ and tend to develop prominent nose cones due to 
the strong magnetic pinching.
This is evident from Fig. (\ref{fig:s1jet}) and (\ref{fig:s2jet}),
where the upper (lower) portion of each gray scale image refers 
to the HLLC (HLL) approximate Riemann solver at $t=100$ and $t=126$, 
respectively. 
The nose cone appears as the extended high pressure region
bounded by the terminal conical shock (mach disk) and the bow shock.
At the Mach disk, the beam is strongly decelerated and magnetic tension 
wards off sideways deflection of shocked material  
into the cocoon. The magnetization parameter, defined as 
$\beta = B_\phi^2/(2\gamma^2p_g)$, is particularly strong in 
the cone above the symmetry axis.   
As pointed out by K99, morphological properties are prevalently 
dictated by the ratio of the kinetic energy flux to the Poynting flux, 
$\kappa = \rho_jh_j\gamma_j^2/B_\phi^2$, which determines 
the magnetization of shocked gas behind the terminal shock.  
An alternative parameter (L05) is the ratio of the
magnetic energy density to the rest mass,
$\sigma = B_\phi^2/(\gamma^2\rho)$:
extended nose cones tend to form for low $\kappa$ (high $\sigma$), 
whereas familiar cocoon structures appear for high values of $\kappa$
(low $\sigma$), regardless of the jet magnetization parameter $\beta$.
Values of $\sigma$ are reported in Table \ref{tab:ic}.
\begin{figure}
 \centerline{\includegraphics[width=115mm]{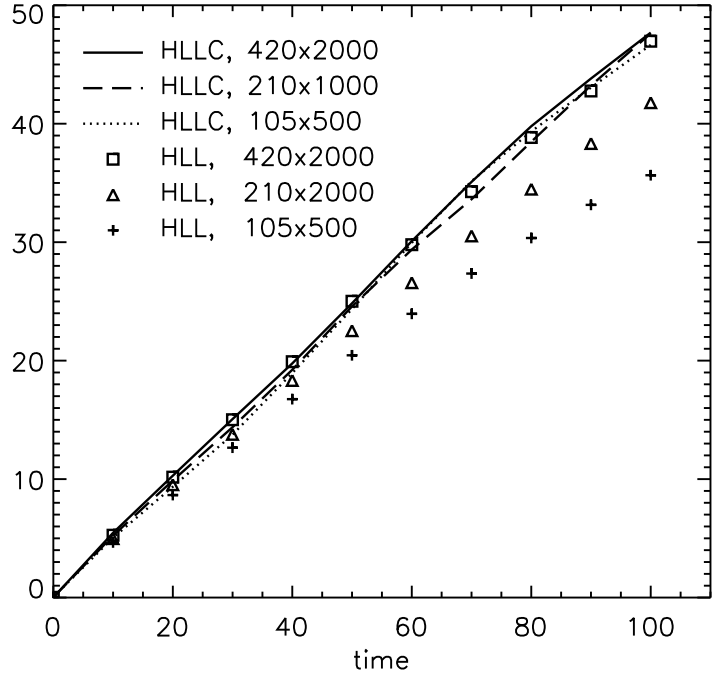}}
 \caption{Jet head positions for model $S1$ computed for 
          three different resolution runs: low ($105\times 500$),
          medium ($210\times1000$) and high ($240\times2000$).
          The HLLC runs are shown using, respectively, dotted, dashed
          and solid lines; for the HLL we use plus signs, triangles and
          squares.}
 \label{fig:ko_pos}
\end{figure}
The positions of the reverse shocks for the HLL and HLLC runs
are, respectively, $z\sim 30$ and $z\sim26$ in model $S1$, 
$z\sim 33$ and $z\sim41$ in model $S2$. 
However, the most striking difference between the HLL and HLLC schemes
is the jet propagation speed obtained from model $S1$.
Indeed, at $t=100$ the head position in the HLLC run is $z \sim 47.6$ to be
compared with a value of $z\sim 41.8$ obtained from the HLL integration.
Note that the final integration time is not the same 
as in K99, where the simulation ends at a later time, $t=110$,
yielding a head position value of $z\sim 46$, apparently more consistent
with the HLL run.
However, in order to further investigate these discrepancies, we carried out an 
additional set of simulations at half and twice the resolution. 
Figure (\ref{fig:ko_pos}) plots the jet head position as function of time 
for the three resolutions in consideration: low ($105\times500$), 
medium ($210\times1000$) and high ($420\times2000$). 
The HLL scheme achieves the same convergence of the HLLC scheme only
at high resolution, whereas the propagation speeds in the HLLC runs
show to be by far less sensitive to the grid size. 

\begin{figure}
 \centerline{\includegraphics[width=115mm]{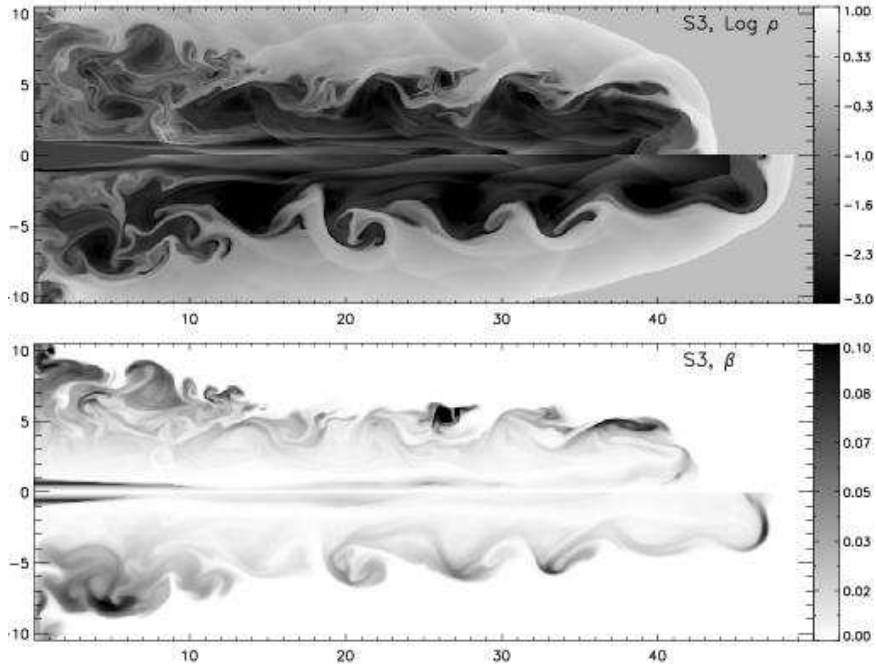}}
 \caption{Model S3 at $t = 126$. The final integration time is the 
          same one used in L05. The meaning of the gray scale images
          in the same as in Fig. (\ref{fig:s1jet}) and (\ref{fig:s2jet}).}
 \label{fig:s3jet}
\end{figure}

This disagreement is remarkably reduced in the second model $S2$, 
where the HLL runs shows a slightly higher propagation speed, but 
again more evident in model $S3$, although with an opposite trend.
Once more we ascribe this behavior to resolution effects, based on the 
fact that the same model in L05 with twice the resolution
(i.e. $40$ zones per beam radius instead of $20$)
shows stronger affinities with the HLLC integration presented here.

When the HLLC solver is employed, the backflow turbulent patterns
in the cocoon are better defined than with the HLL scheme, where
the larger numerical viscosity inhibits the formation of small
scale structure at lower resolution.
This is corroborated by a simple qualitative comparison with the results 
of L05, who used twice the resolution employed here and the HLL solver. 
The small scale structural details provided by the HLLC 
approach appear to be almost equivalent even at half the resolution.
This consideration together with the convergence test previously shown 
confirms, indeed, that the ability to describe tangential 
and shear waves largely improves the quality of solution, particularly
in the multidimensional case.
  
Finally, we remind that the computational cost of the HLLC scheme over the 
traditional HLL Riemann solver is less than $7\%$ and we believe that this 
largely justifies its use in  Godunov-type codes.

\section{Discussion}\label{sec:discussion}
%
%
%

Numerical simulations of relativistic magnetized jets 
with toroidal magnetic fields have been presented. 
The three models considered show formation of conspicuous nose
cones and jet confinement with increasing magnetization.
The morphological properties favorably compare with the 
results of previous investigators. 
Simulations have been conducted by using the recent relativistic 
hydrodynamics HLLC solver by Mignone \& Bodo, opportunely 
extended to the special case of vanishing normal 
component of magnetic field.
In order to assess the validity of the new algorithm, 
simulations have been directly compared with the 
traditional HLL Riemann solver \cite{dZBL03}.
It is found that the HLLC approach largely improves the quality 
of solution over the HLL scheme by considerably reducing the amount
of numerical viscosity.
This is manifested in a sharper resolution of fine scale 
structures in proximity of shear and tangential flows. 
Furthermore, convergence tests at different resolutions 
have shown that, with the HLLC scheme, the jet head position 
has little dependence on the grid size. This is a remarkable
property for a shock-capturing scheme and has to be consolidated through 
further numerical testing.
Generalization to the full relativistic MHD case is underway.

\bibliographystyle{klunamed}
{}

\end{article}
\end{document}